\documentclass[aps,twocolumn,showpacs]{revtex4}
\usepackage{graphics}
\usepackage{graphicx}

\begin{document}
\title{Saturation of interband absorption in graphene}
\author{F.T. Vasko}
\email{ftvasko@yahoo.com}
\affiliation{Institute of Semiconductor Physics, NAS of Ukraine,
Pr. Nauky 41, Kiev, 03028, Ukraine}
\date{\today}

\begin{abstract}
The transient response of an intrinsic graphene, which is caused by the ultrafast
interband transitions, is studied theoretically for the range of pumping correspondent
to the saturated absorption regime. Spectral and temporal dependencies of the
photoexcited concentration as well as the transmission and relitive absotption
coefficients are considered for mid-IR and visible (or near-IR) spectral regions at
different durations of pulse and broadening energies. The characteristic intencities
of saturation are calculated and the results are compared with the experimental data
measured for the near-IR lasers with a saturable absorber. The negative absorption
of a probe radiation during cascade emission of optical phonons is obtained.
\end{abstract}

\maketitle

\section{Introduction}
The character of nonlinear response under the ultrafast interband excitation of
an intrinsic graphene is determined by a several physical processes which are
dependent on conditions of pumping. Under a low excitation level, when the
one-photon transitions take place, the energy relaxation and recombination of
photoexcited carriers were studied with the use of the time-resolved pump-probe
measurements, see experimental data and theoretical discussions in Refs. 1 and 2,
respectively. Under an extremely high pumping, the multi-quantum transitions,
which cause the harmonics generation and the hybridization of electron-hole states,
take place. This regime is not investigated completely for graphene, see general
consideration for bulk materials or quantum wells in Sects. 10 and 56 of Ref. 3. The
nonlinear response is also possible within the single-photon approach because the
Rabi oscillations of coherent response should take place in graphene if a pulse
duration is less 100 fs. \cite{4} Beside of this, the nonlinear regime of energy
relaxation and recombination due to the Pauli blocking effect takes place if the
photogeneration rate is comparable to the energy relaxation or recombination
rates. A saturation of transient absorption, which have been investigated recently,
\cite{5,6} is the most important manifistation of such a regime. It is because this
phenomenon was exploited for realization of an ultrafast laser with a graphene
saturable absorber in the telecommunication spectral region. To the best of our
knowledge, a complete theoretical treatment of the saturation mechanism is not
performed yet and an investigation of this phenomena is timely now.

In this paper, we consider the temporal nonlinear evolution of carriers under
photoexcitation by the ultrafast pulses in mid-IR and visible (or near-IR) spectral
regions. For the mid-IR pump, one can neglect the relaxation and recombination
processes (the quasielastic relaxation due to acoustic phonons remains uneffective
up to nanoseconds) and the  transient distribution of nonequilibrium carriers is
determined by the broadening of interband transitions due to an elastic scattering
and by the parameters of excitation. For the excitation energies, $\hbar\Omega$, above
the optical phonon energy, $\hbar\omega_\eta$ ($\eta =\Gamma ,~K$ labels the phonon
modes correspondent to the intra- and intervalley transitions), when a cascade
emission of the optical phonons dominates in relaxation, the transient distribution transforms into a set of peaks. The effective electron-hole recombination takes place
if the lowest peak is placed around the half-energy of optical phonon,
$\hbar\omega_\eta /2$. So that, the character of response modifies essentially if
the frequency $\Omega$ varies over the $\omega_\eta$-range.

The saturation process is described within the framework of the temporally local
approach, when the decoherentization time (which determines the broadening of the
interband transitions) is shorter in comparision with the duration of pumping.
Spectral and temporal dependencies of the photoexcited concentration and the response
on the probe radiation (transmission and absorption coefficients) are presented.
The thresholds for saturation of response are estimated to be about 0.2 MW/cm$^2$,
60 MW/cm$^2$, and 0.6 GW/cm$^2$ for $\hbar\Omega\sim$0.12, 0.8, and 1.5 eV,
respectively (mid-IR, near-IR, and visible spectral regions). These results are
dependent on the decoherentization and relaxation mechanisms and the electrodynamics
conditions. Their are discussed in comparison with the experimental data for the
near-IR pumping case \cite{5,6}. Conditions for the transient negative absorption
of a probe radiation during cascade emission of optical phonons are also analyzed
(this phenomenon under a steady-state pumping was considered recently \cite{7} in
connection with a possibility of the THz lasing effect).

The consideration below is organized as follows. The temporally local approach
for description of the response of photoexcited carriers is developed in Sec. II.
Spectral and temporal dependencies of the response are described in Sects. III and
IV for the cases of excitation in mid-IR and visible (or near-IR) spectral regions, respectively. A discussion of experimental data, the list of assumptions used, and concluding remarks are given in the last section. In Appendix we consider the
mechanism of saturation caused by the collisionless Rabi oscillations.

\section{Temporally local approach}
Since the symmetry of the energy spectrum and scattering processes for electrons
and holes in an intrinsic graphene, we describe the phenomena under consideration
by the same distribution functions for the both types of carriers, $f_{pt}$.
According to Refs. 2b and 4, the kinetic equation for $f_{pt}$ takes form:
\begin{equation}
\frac{df_{pt}}{dt} =\nu_{pt}(1-2f_{pt})+J(f_t|p)
\end{equation}
and it should be solved with the initial condition $f_{pt\to\infty}=0$. Here
$J(f_t|p)$ is the collision integral, which is described the relaxation and
recombination processes, and $\nu_{pt}$ is the interband generation rate due to
the in-plane electric field ${\bf E}w_t\exp (-i\Omega t)+c.c.$, where ${\bf E}$ is
the field strength, $\Omega$ is the pumping frequency, and $w_t$ is the envelope
form-factor of pulse with duration 2$\tau_p$ centered at $t=0$. Supposing that
$\tau_p$ exceeds the dephasing time, we have used in Eq. (1) the temporally local
approach with the separated filling factor, $(1-2f_{pt})$, and with the rate of photoexcitaion:
\begin{equation}
\nu_{pt}=\nu_R w_t^2\Delta\left(\frac{2\upsilon p-\hbar\Omega}{\gamma} \right), ~~~~ \nu_R=\frac{\pi (eE\upsilon /\Omega )^2}{\hbar \gamma} .
\end{equation}
Here $\upsilon =10^8$ cm/s is the velocity of neutrinolike quasiparticles, $2\gamma$
is the broadening of the interband excitation described by the phenomenological
factor $\Delta (z)$. Below we consider the Lorentzian lineshape of photoexcitation,
when
$\Delta (z)=\left[\pi (1+z^2)\right]^{-1}$ and the Gaussian temporal envelope
$w_\tau =\sqrt[4]{2/\pi}\exp [-(t/\tau_p)^2]$.

The solution of Eq. (1) determines both the photoinduced concentration, which
given by the standard formula
\begin{equation}
n_t =\frac{2}{\pi\hbar^2}\int\limits_0^\infty {dpp}f_{pt},
\end{equation}
and the transient response on a probe radiation of frequency $\omega$ ($\propto\exp
(-i\omega t)$, which is described by the dynamic conductivity $\sigma_{\omega t}$.
For the collisionless case $\hbar\omega /\gamma\gg 1$, when the parametric dependency
on time takes place, \cite{9} the real part of $\sigma_{\omega t}$ is written as
follows:
\begin{equation}
{\rm Re}\sigma_{\omega t}=\frac{e^2}{4\hbar}(1-2f_{p_\omega ,t}) ,
\end{equation}
where $p_\omega =\hbar\omega /\upsilon$. The imaginary part of $\sigma_{\omega t}$
is determined through ${\rm Re}\sigma_{\omega t}$ with the use of the dispersion
relation and one can check that the carrier-induced contribution to ${\rm Im}
\sigma_{\omega t}$ appears to be weak in comparison with (4) for the peak-like
distributions of carriers considered below. Thus, the only filling factor in ${\rm Re}
\sigma_{\omega t}$ is responsible for the nonlinear behavior of the response
under consideration.

We restrict ourselves by the the geometry of normal propagation of radiation.
The relative absorption of graphene sheet, $\xi_{\omega t}$, as well as the
reflection and transmission coefficients, $R_{\omega t}$ and $T_{\omega t}$,
are determined through $\sigma_{\omega t}$. Since the energy conservation
requirement, \cite{8}
\begin{equation}
R_{\omega t}+T_{\omega t}+\xi_{\omega t}=1 ,
\end{equation}
we consider below only the absorption and transmission coefficients:
\begin{eqnarray}
\xi_{\omega t}\simeq\frac{16\pi}{\sqrt\varepsilon c}\frac{{\rm Re}\sigma_{\omega t}}
{|1+\sqrt\epsilon +4\pi\sigma_{\omega t}/c|^2}\approx\xi_m (1-2f_{p_\omega t}) ,
\nonumber \\
T_{\omega t}\simeq\frac{4\sqrt \epsilon}{\left| 1+\sqrt\epsilon +4\pi
\sigma_{\omega t}/c\right|^2}\approx\frac{T_m}{(1-af_{p_\omega t})^2} .
\end{eqnarray}
Here $\sqrt\epsilon$ is the refraction index of a thick substrate (for SiO$_2$
substrate $\sqrt\epsilon\simeq$1.45 and dispersion of $\epsilon$ can be neglected)
and we approximately separated the carrier contributions using the coefficients
$\xi_m\approx 4\pi\alpha /\left[\sqrt{\epsilon}(1+\sqrt{\epsilon})\right]$,
$T_m\approx 4\sqrt{\epsilon}/(1+\sqrt{\epsilon}+\pi\alpha )^2$, and $a\approx
2\pi\alpha /(1+\sqrt{\epsilon}+\pi\alpha )$ with $\alpha =e^2/\hbar c$. Notice,
that the negative absorption regime $\xi_{\omega t}<0$ takes place if
$f_{p_\omega ,t}>1/2$, under the population inversion condition (see discussion
in Sec. IV).

At $\omega =\Omega$ these relations describe the propagation of pumping pulse
with the time-dependent intensity $Sw_t^2$, where $S$ is the maximal intensity. Performing
the averaging of (6) over the pulse duration one obtains
\begin{equation}
\left|\begin{array}{*{20}c} \xi_S \\ T_S \end{array}\right| =
\int\limits_{-\infty}^\infty\frac{dt}{\tau_p}w_t^2\left|\begin{array}{*{20}c}
\xi_{\Omega t} \\ T_{\Omega t} \end{array} \right| ,
\end{equation}
where we used $\int\limits_{-\infty}^\infty dtw_t^2/\tau_p =1$. Below we
solve Eq. (1) and analyze the responses (6) and (7) for different parameters
of pump and probe radiations.

\section{Mid-IR excitation}
We consider here the mid-IR pumping case when the energy relaxation of carriers
is ineffective and $J(f_t|p)$ in Eq. (1) can be neglected. As a result, the
solution of the problem (1) takes form:
\begin{equation}
f_{pt}=\int\limits_{-\infty}^t dt'\nu _{pt'}\exp\left( -2\int\limits_{t'}^t
d\tau\nu_{p\tau}\right) .
\end{equation}
Evolution of such a distribution from zero value at $t\ll -\tau_p$ to the
saturated peak with the maximal value $f_{max}=1/2$ is shown in Fig. 1a versus
dimensionless time and energy at the pumping intensity $S=$1 MW/cm$^2$. Temporal
dependencies of $f_{p_\Omega t}$ at different $S$ are shown in Fig. 1b. These
calculations were performed for $\hbar\Omega\simeq$120 meV (pumping by CO$_2$-laser),
the pulse duration $2\tau_p\simeq$1 ps, and the broadening energy $\gamma\simeq$6
meV which is in agreement with the mobility data for the case of elastic
scattering. \cite{9} The temporally-dependent photoinduced concentration
$n_t$ is plotted Fig. 1c for the same parameters. The saturated concentration
versus intensity, which is attained at $t>\tau_p$, is presented for $\gamma =$6
and 12 meV in Fig. 1d. These dependencies can be fitted as
\begin{equation}
n_S\approx\frac{bS}{1+S/S_n} ,
\end{equation}
where $b\simeq$6 or 12.2 MW$^{-1}$ [$n_S$ is measured in 10$^{11}$ cm$^{-2}$] and $S_n\simeq$1.76 or 10 MW/cm$^2$ for and $\gamma =$6 or 12 meV,
respectively.
\begin{figure}[ht]
\begin{center}
\includegraphics{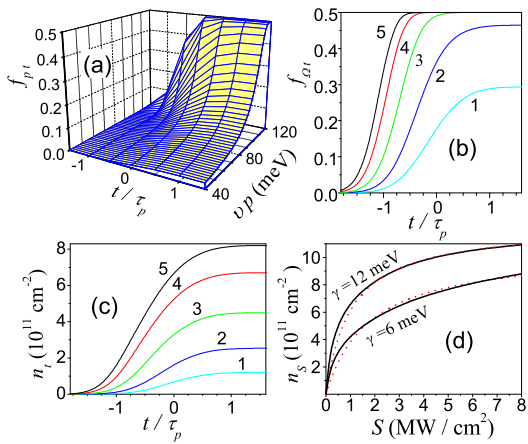}
\end{center}\addvspace{-1 cm}
\caption{(Color online) (a) Photoexcited distribution $f_{pt}$ versus energy
$\upsilon p$ and dimensionless time, $t/\tau_p$ at mid-IR pumping level $S=$1
MW/cm$^2$. (b) Temporal evolution of $f_{\Omega t}\equiv f_{p_\Omega t}$ at
$S=$0.1, 0.3, 1, 3, and 6 MW/cm$^2$ (marked as 1-5). (c) Potoinduced concentration
versus $t/\tau_p$ for the same conditions as in panel (b). (d) Concentration
$n_S$ at $t/\tau_p\to\infty$ versus $S$ for the different broadening
energies $\gamma$. Dotted curves are correspondent to the fit (9). }
\end{figure}

The relative absorption and transmission coefficients of a probe radiation
of frequency $\omega$ are determined through $f_{p_\omega t}$ according to Eqs. (6). Spectral and temporal dependencies of $\xi_{\omega t}$ are shown in Fig. 2a for the conditions used in Fig. 1a. Since $af_{p_\omega t}\ll 1$, the peak of relative
transmission $T_{\omega t}/T_m$ resembles $f_{p_\Omega t}$ presented in Fig. 1a.
Here $T_m= \simeq$0.95 is the transmission coefficient without for non-doped
graphene. The temporally-dependent relative absorption and transmission at the pumping
frequency $\Omega$ and at different $S$ are presented in Figs. 2b and 2c, respectively.
The saturated values of $\xi_{S}/\xi_m$ and $T_{S}$ versus intensity are plotted
in the upper and lower panels of Fig. 2d for the parameters used in Fig. 1d ($\xi_{S}$
and $T_S$ have only a weak dependency on $\gamma$). These curves can be fitted as
\begin{equation}
\xi_S\approx\frac{\xi_m} {1+S/\overline{S}}, ~~~~
T_S\approx T_m+\frac{hS}{1+S/\overline{S}} ,
\end{equation}
where $\overline{S}\approx$0.2 MW/cm$^2$ and $h\approx$0.09 cm$^2$/MW. The saturation
of $\xi_{S}$ and $T_S$ takes place at lower threshold intensities in comparision
to $n_S$, c. f. Figs. 1c and 2d, 2e. Thus, for the pumping range
$\geq$1 MW/cm$^2$ the one-photon absorption is suppressed and a damage of graphene
by mid-IR radiation with $\tau_p\lesssim$1 ps is not possible.
\begin{figure}[ht]
\begin{center}
\includegraphics{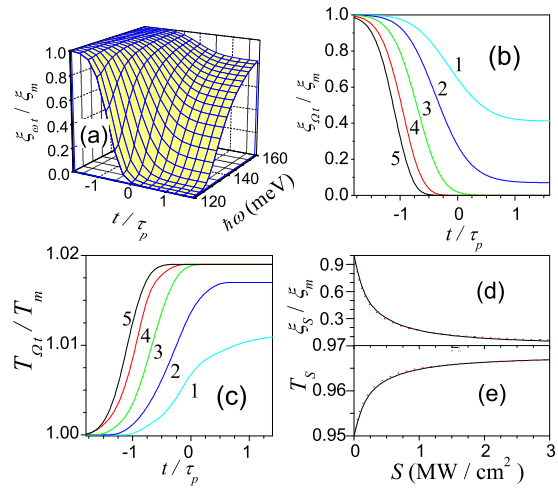}
\end{center}\addvspace{-1 cm}
\caption{(Color online) (a) Spectral and temporal dependencies of relative
absorption, $\xi_{\omega t}$ at $S=$1 MW/cm$^2$. (b) Temporal evolution
of $\xi_{\Omega t}$ at $S$ used in  Fig. 1b (marked). (c) The same as in panel
(b) for transmission, $T_{\Omega t}$. (d) Avaraged over pulse absorption and
transmission coefficients (upper and lower panels, respectively) versus $S$
[dashed curves are correspondent to Eq. (10)]. }
\end{figure}

\section{Cascade emission effect}
In this section we consider the photoexcitation by visible and near-IR radiation, when
the cascade emission of optical phonons should be taken into account in Eq. (1).
For the temperatures below the optical phonon energies, the spontaneous emission
processes are only essential and the collision integral is given by the
finite-difference form (see evaluation in Refs. 2b and 10)
\begin{eqnarray}
J\left( f_t |p\right)=\sum_\eta\left[\nu_{p+p_\eta}\left( 1-f_{pt}\right)f_{p+
p_\eta t} \right. \\
\left. -\nu_{p-p_\eta}\left( 1-f_{p-p_\eta t}\right)f_{pt}-\widetilde\nu_{p_\eta-p}
f_{p_\eta -pt} f_{pt}\right] . \nonumber
\end{eqnarray}
Here $\eta =\Gamma ,~K$ is correspondent to the intra- and intervalley transitions
with the energy transfer, $\hbar\omega_\eta$, and the momentum transfer, $p_\eta =\hbar\omega_\eta /\upsilon$. The last contribution of Eq. (11) is responsible for
the recombination process while the first and second terms describe the interband
cascade relaxation of carriers. The relaxation rates $\nu_p$ and $\widetilde\nu_p$ are proportional to the density of states, $\nu_p\approx\widetilde{\nu}_p\approx\theta (p)\upsilon_\eta p/\hbar$, where the characteristic velocities $\upsilon_{\Gamma ,K}$
can be estimated crudely as $\upsilon_K\approx 2\times 10^6$ cm/s and $\upsilon_{\Gamma}\approx 10^6$ cm/s. \cite{10} Thus, the $K$-mode emission gives a dominant contribution
to the relaxation process; moreover, the only interband recombination is possible
in the passive region, $0<\upsilon p<\hbar\omega_K =$170 meV. Below we neglect other
relaxation processes, so that a peak-like transient distribution of carriers takes
place due to the negligible phonon dispersion and a narrow distribution of photoexcited carriers, under the condition $\gamma\ll\hbar\omega_K$. For the sake of simplicity,
the cases of effective or suppressed recombination, when the lower peak in the passive region is placed around or outside the energy $\hbar\omega_K$ are considered. It is convenient to analyze calculations for the near-IR and visible pumping cases separately.
\begin{figure}[ht]
\begin{center}
\includegraphics{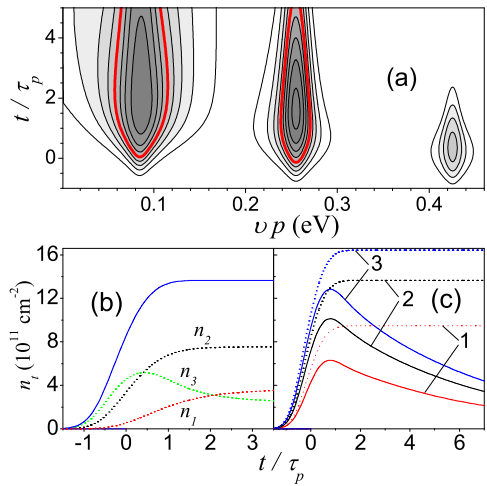}
\end{center}\addvspace{-1 cm}
\caption{(Color online) (a) Contour plots of photoexcited distributions $f_{pt}$
versus energy and time for pulse duration $2\tau_p=$0.6 ps at pumping level 200
MW/cm$^2$ and $\hbar\Omega =$850 meV. (b) Transient evolution of concentration
$n_t$ and and populations of peaks around $\sim$43, 213, and 383 meV (marked as
$n_1$, $n_2$, and $n_3$, respectively), for the same conditions as in panel (a).
(c) Evolution of $n_t$ for pumping levels $S=$100, 200, and 300 MW/cm$^2$ (marked
as 1, 2, and 3). Solid and dashed curves are plotted for $\hbar\Omega =$ 850 meV
and 765 meV. }
\end{figure}

\subsection{Near-IR pumping}
First, we consider the three-step cascade processes under the near-IR pumping with wavelengths around $\sim 1.5~\mu$m and the pulse duration determined by $\tau_p =$
0.3 ps. We consider the regimes of the enhanced or suppressed recombination supposing $\hbar\Omega =$ 850 meV or 765 meV. For this energy region, the broadening of
photoexcited peak is taken as $\gamma\simeq$18 meV, so that $\hbar /\gamma\ll\tau_p$.
The numerical solution of Eq. (1) with the collision integral (11) is performed
with the use of the temporal iterations \cite{11} at different $S$. Figure
2a shows the contour plot of the three-peak distribution function $f_{pt}$ for the
case of efficient recombination ($\hbar\Omega =$ 850 meV) at $S=$200 MW/cm$^2$.
The carrier concentrations over the peaks 1-3 and the total consentration $n_t$
given by Eq. (3) are shown in Fig. 3b for the same parameters as in Fig. 1a. Since
the relaxation rate in Eq. (11) is proportional to the density of states, $\nu_p
\propto p$, the bottleneck effect takes place under the transition between the second
and third peaks and $n_2$ exceeds $n_{1,3}$. The transient evolution of concentration
for different $S$ is shown in Fig. 3c where the maximal concentration exceeds
10$^{12}$ cm$^{-2}$ at $t\sim\tau_p$ and $S\geq$0.3 GW/cm$^2$. During the further
evolution, $n_t$ decays due to the recombination process. The case of the suppressed recombination ($\hbar\Omega =$765 meV) is different because of, first, the peaks
are shifted below (about 43 meV) and, second, the decreasing of $f_{pt}$ and $n_t$
due to recombination is absent. The saturated concentrations (dashed curves in
Fig. 3c) exceed the peak concentrations (solid curves in Fig. 3c) by factor
$\sim$1.3. Note, that $n_t$ decreases with increasing of $\gamma$ at fixed $S$
(not shown in Fig. 3).
\begin{figure}[ht]
\begin{center}
\includegraphics{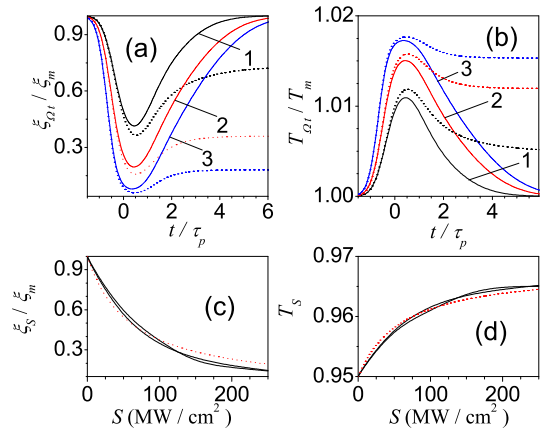}
\end{center}\addvspace{-1 cm}
\caption{(Color online) (a) Transient evolution of relative absorption for
$\hbar\Omega =$ 850 and 765 meV (solid and dotted curves, respectively) at
$S=$50, 100 and 200 MW/cm$^2$ (marked as 1-3).
(b) The same as in panel (a) for transmission coefficient. (c) Avaraged over
pulse absorption versus $S$ [dashed curve is correspondent to Eq. (9)].
(d) The same as in panel (c) for transmission coefficient. }
\end{figure}

Transient evolutions of the absorption and transmission coefficients given by Eq. (6)
are shown in Figs. 4a and 4b at the different pumping frequencies $\Omega$ (solid and
dashed curves) and at different $S$. For $t/\tau_p\leq$0.5, the temporal evolution of
$\xi_{\Omega t}$ and $T_{\Omega t}$ do not dependent on the character of recombination.
For $t/\tau_p\lesssim$1.5 this evolution is completely different: a quenching of photoresponse or a steady-state contribution take place for the effective or suppressed recombination cases. At $S\geq$300 MW/cm$^2$ and $t/\tau_p\sim$0 one obtains the
saturated absorption around $\xi_{\Omega t}\sim$0.1. The negative absorption takes
place for a probe radiation with $\hbar\omega$ around the first and second peaks. It
is because $f_{p_{\omega}t}>1/2$, see Eq. (4) and the contour plot in Fig. 3, where
the regions of negative absorption are separated by the thick (red) curves. Thus, the negative absorption (and a possible stimulated emission of mid-IR radiation) is
realized at $S\geq$100 MW/cm$^2$ during time intervals $t\lesssim 5\tau_p$.

The absorption and transition coefficients averaged over pulse duration according to
Eq. (7) are shown in Figs. 4c and 4d.  Since the transient response at
$|t|\lesssim\tau_p$ does not depend on the recombination mechanism (see Figs. 3c, 4a,
and 4b), the variation of $\xi_S$ and $T_S$ with $\hbar\Omega$ is less than 5\%. These dependencies can be fitted by Eq. (10) with the characteristic intensity $\overline{S}\approx$60 MW/cm$^2$ and the coefficient $h\approx$0.3 cm$^2$/GW.

\begin{figure}[ht]
\begin{center}
\includegraphics{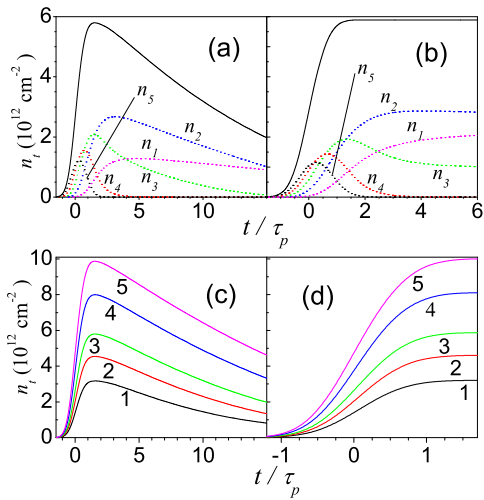}
\end{center}\addvspace{-1 cm}
\caption{(Color online) (a) Transient evolution of concentration $n_t$ and and
populations of peaks around $\sim$0.09, 0.26, 0.43, 0.6, and 0.77 eV (marked
as $n_{1-5}$, respectively) for pulse duration $2\tau_p=$0.4 ps at pumping level
0.4 GW/cm$^2$ and $\hbar\Omega =$1.53 eV. (b) The same as in panel (a) at
$\hbar\Omega =$1.615 eV for peak's positions $\sim$0.13, 0.3, 0.47, 0.64,
and 0.81 eV marked as $n_{1-5}$. (c) Evolution of $n_t$ for pumping levels $S=$0.2,
0.3, 0.4, 0.6 and 0.8 GW/cm$^2$ (marked as 1-5, respectively) for
$\hbar\Omega =$1.53 eV. (d) The same as in panel (c) for $\hbar\Omega =$1.615 eV. }
\end{figure}

\subsection{Visible pumping}
Next, we consider the visible light pumping, with wavelengths around
$\sim 0.75~\mu$m, using the pulse duration $2\tau_p =$0.4 ps and the broadening
$\gamma\approx$34 meV (so that $\hbar /\gamma\ll\tau_p$). Supposing $\hbar\Omega =$
1.53 and 1.615 eV for the enhanced and suppressed recombination regimes one arrive
to the distribution function formed during the five-step cascade process. Transient
evolutions of the concentrations over the peaks 1-5 and of the total concentration
$n_t$ at $S=$0.4 GW/cm$^2$ are shown in Figs. 5a and 5b for the cases of enhanced and suppressed recombination, respectively. Similarly to the near-IR pumping case, the
upper peak concentrations decrease fast at $t>\tau_p$ and a maximal population of
the second peak takes place due to the bottleneck effect. Once again, at $t<\tau_p$
the shapes of $n_t$ are the same for the both cases. At $t>\tau_p$ a quenching of
$n_t$ due to recombination takes place in Fig. 5a while there is no a decreasing
of $n_t$ in Fig. 5b. The temporal dependencies of concentration for different $S$
are shown in Figs. 5c and 5d for the two recombination regimes under consideration.
The maximal concentration range up to 10$^{13}$ cm$^{-2}$ at $t\sim\tau_p$ and
$S\approx$1 GW/cm$^2$.
\begin{figure}[ht]
\begin{center}
\includegraphics{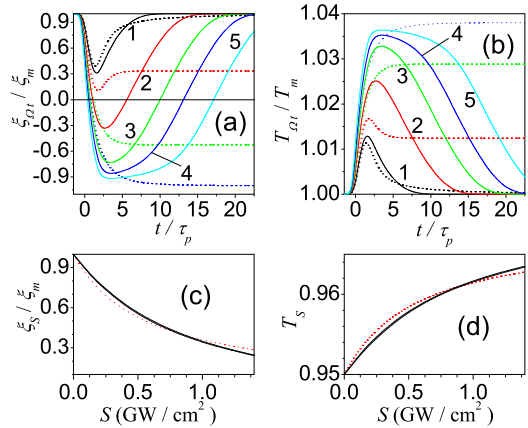}
\end{center}\addvspace{-1 cm}
\caption{(Color online) (a) Transient evolution of relative absorption for
$\hbar\Omega =$1.53 and 1.615 eV, (solid and dotted curves respectively) at $S=$0.2,
0.4, 0.6, 0.8 and 1.2 GW/cm$^2$ (marked as 1-5). (b) The same as in panel (a) for transmission coefficient. (c) Avaraged over pulse absorption versus $S$ [dashed curve is
correspondent to Eq. (9)]. (d) The same as in panel (c) for transmission coefficient. }
\end{figure}

The temporal evolution of $\xi_{\Omega t}$ and $T_{\Omega t}$ at frequency $\Omega$
(solid and dotted curves are correspondent to the two recombination cases under
consideration) are plotted in Figs. 5a and 5b. By analogy with Sect. IVA, the negative absorption regime takes place at $S\geq$0.3 GW/cm$^2$. Beside of this, the conditions
$\xi_{\omega t}<0$ take place around the peak positions at $\omega <\Omega$ (not plotted, see a similar behavior in Fig. 3a); for the first and second peaks the
negative absorption regime is realized up to $t\sim 5\tau_p$ at $S\geq$0.1 GW/cm$^2$.
In addition, at $t<0.5\tau_p$ the response does not dependent on recombination and
at $t>1.5\tau_p$ a damping or time-independent response is realized for the effective
or suppressed recombination.

The averaged according to Eq. (7) absorption and transmission coefficients, which
do not depend on the recombination mechanism, are plotted in Figs. 6c and 6d. Once
again, $\xi_S$ and $T_S$ can be fitted by Eqs. (10) with the characteristic intensity $\overline{S}\approx$0.56 GW/cm$^2$ and the coefficient $h\approx$0.03 cm$^2$/GW. Since
the departure rate from the photoexcited peak increases if $\hbar\Omega$ grows, the characteristic intensity $\overline{S}$ is also increased in the visible spectral
region in comparison with the near-IR pumping case.

\section{Discussion and conclusions}
To summarize, we have developed the nonlinear theory of transient response of an intrinsic graphene under the ultrafast interband excitation. Within the local time
approach, the conditions of saturation of absorption were found in the mid-IR, near-IR,
and visible spectral regions. In addition, we have demonstrated a possibility for the
stimulated mid-IR radiation due to the bottleneck effect during the cascade emission of
optical phonons. Our consideration is based on the set of assumptions about relaxation mechanisms. First of all, the phenomenological model for the broadening with the characteristic energy $\gamma$ is used for description of the intersubband transitions.
In Sects. III and IV we estimated $\gamma$ from the experimental data for the departure
relaxation rates. \cite{1,2,9} Secondary, a simplified description of energy
relaxation is employed. We neglect the Coulomb scattering which is not a dominant
relaxation channel at $t\lesssim\tau_p$, so that the results for $\xi_S$ and $T_S$
should not be modified essentially. But a transient distribution at $t\gg\tau_p$
and a condition for the negative absorption of a probe radiation in the mid-IR region
can be modified. Also, a possible contribution of the substrate vibration \cite{12}
is not taken into account. These points require a special consideration but, anyway,
our calculation gives a lower bound of $S$. The other assumptions (parameters for
the electron-phonon coupling, conditions for the temporally-local approach, and
description of the interband response) are rather standard for the calculations
of the optical properties and the relaxation phenomena in graphene. In addition,
an inhomogenity of pumping, which causes the lateral diffusion of carriers, \cite{13}
and a heating of phonons \cite{14} may be essential; these phenomena requre a special treatment, both experimental and theoretical.

We turn now to discussion of the experimental data available for the near-IR spectral
region. \cite{5,6} Numerical estimates for the saturation thresholds and for the
concentrations of the photoexcited carriers are in a qualitative agreement with the
consideration performed. But an accurate comparison with the results presented is not
possible for the two reasons. First, the graphene structure was embedded into the
laser cavity in \cite{5,6} so that the electrodynamical conditions (for a propagated,
reflected, and absorbed radiation) were different from the simple geometry considered
here. Second, the multi-layer graphene or the graphene flakes were used, while a
single-layer graphene case was not under a detailed treatment. Thus, a special
measurements with the use of the simplest geometry of a well-characterized sample
placed over a semi-infinite substrate are necessary.

In closing, we have analyzed theoretically the conditions for realization of an
efficient graphene-based saturable absorber and have performed a comparison with
the experimental data. More extended treatment of this phenomena under near-IR
pumping, including an above-mentioned special measurements, in order to improve an efficiency of the graphene based saturable absorber in the lasers for telecommunications.
An additional study in the mid-IR and visible spectral regions should be useful
for verification of different relaxation mechanisms. \\

The author would like to thank E. I. Karp for insightful comments.

\appendix*
\section{Rabi oscillations regime}
Below we describe the saturation of the averaged absorption and transmission
coefficients (7) under an ultrafast pumping for the case when the Rabi oscillations conditions are satisfied. \cite{4} The collisionless regime of response is
described by the $S$-dependent contribution to the distribution function
\begin{equation}
1-2f_{pt}=\cos\left(\sqrt{\frac{S}{S_R}}\int\limits_{-\infty}^t
\frac{dt'}{\tau _p}w_{t'}\right) .
\end{equation}
Here the characteristic intensity is given by
\begin{equation}
S_R=\frac{\sqrt{\epsilon}c}{4\pi }\left(\frac{\hbar\Omega}{e\tau_p\upsilon}
\right)^2
\end{equation}
and $S_R\simeq$0.6 MW/cm$^2$ for CO$_2$ pumping with $\tau_p\simeq$0.1 ps.
For the near-IR or visible pumping with $\tau_p\simeq$30 fs, one obtains
$S_R\simeq$0.3 or 1 GW/cm$^2$. Notice, that $S_R\propto (\Omega /\tau_p)^2$
and (A.1) is not dependent on any other parameter if $\tau_p$ is shorter than
the dephasing relaxation time.
\begin{figure}[ht]
\begin{center}
\includegraphics{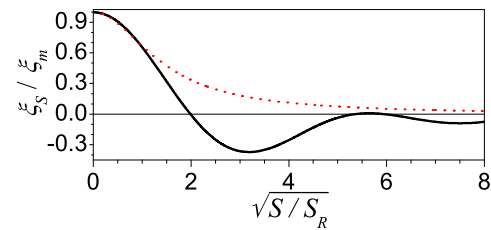}
\end{center}\addvspace{-1 cm}
\caption{(Color online) Normalized absorption coefficient given by (A.3) versus dimensionless intensity $S/S_R$. Dotted curve presents a monotonic fit. }
\end{figure}

Substituting the distribution (A.1) into Eqs. (6) and (7) one obtains the following
analytical expression for the relative absorption
\begin{equation}
\frac{\xi_S}{\xi_m}=\int\limits_{-\infty}^\infty\frac{dt}{\tau_p} w_t^2 \cos
\left(\sqrt{\frac{S}{S_R}}\int\limits_{-\infty}^t\frac{dt'}{\tau _p}w_{t'}
\right)
\end{equation}
while the transmission coefficient is given by $T_S\approx T_m (1-\widetilde{a}\xi_S )$
with $\widetilde{a}=\sqrt{\epsilon}(1+\sqrt{\epsilon})/2$. In Fig. 7 we plot the
function $\xi_S /\xi_m$ versus dimensionless intensity $S/S_R$ and the oscillating
character of response at $S/S_R\geq$4. The oscillations appears due to the dynamic
inversion of transient population, see Ref. 4. The fit of (A.3) at $S/S_R\leq 1$ is
given by Eq. (10) with the characteristic intensity $\overline{S}=2S_R$.

\end{document}